\documentclass[prb,amsfonts,amssymb,floats,twocolumn,showpacs,aps]{revtex4}

\usepackage[final,dvips]{epsfig}
\usepackage{amsmath}

\newcommand{\w}{\omega}

\newcommand{\Tsat}{T_{\rm sat}}
\newcommand{\Tcut}{T_{\rm cut}}
\newcommand{\ncut}{n_{\rm cut}}
\newcommand{\mcut}{m_{\rm cut}}

\begin{document}

\title[]
{
Numerical Renormalization Group for the sub-ohmic spin-boson model:\\
A conspiracy of errors
}
\author{Matthias Vojta}
\affiliation{Institut f\"ur Theoretische Physik,
Technische Universit\"at Dresden, 01062 Dresden, Germany}

\date{\today}

\begin{abstract}
The application of Wilson's Numerical Renormalization Group (NRG) method to dissipative
quantum impurity models, in particular the sub-ohmic spin-boson model, has led to
conclusions regarding the quantum critical behavior which are in disagreement with those
from other methods and which are by now recognized as erroneous. The errors of NRG
remained initially undetected because NRG delivered an internally consistent set of
critical exponents satisfying hyperscaling. Here we discuss how the conspiracy of two
errors -- the Hilbert-space truncation error and the mass-flow error -- could lead to
this consistent set of exponents. Remarkably, both errors, albeit of different origin,
force the system to obey naive scaling laws even when the physical model violates naive
scaling. In particular, we show that a combination of the Hilbert-space truncation and
mass-flow errors induce an artificial non-analytic term in the Landau expansion of the
free energy which dominates the critical behavior for bath exponents $s<1/2$.
\end{abstract}


\maketitle


\section{Introduction}

The Numerical Renormalization Group method,\cite{wilson75,nrgrev} originally
developed\cite{wilson75} by Wilson for the Kondo model and subsequently applied to a
variety of impurity models with fermionic baths, was generalized in
Refs.~\onlinecite{BTV,BLTV} to the case of bosonic baths with an eye towards dissipative
impurity models. In particular, it has been applied to the spin-boson
model,\cite{leggett}
\begin{equation}
{\cal H}_{\rm SB}=-\frac{\Omega}{2}\sigma_{x} + \frac{\epsilon}{2}\sigma_{z}
+\frac{\sigma_{z}}{2} \sum_{i}
    \lambda_{i}( b_{i} + b_{i}^{\dagger} )
+\sum_{i} \omega_{i} b_{i}^{\dagger} b_{i}
\label{sbm}
\end{equation}
where $\sigma_z=\pm 1$ are the local impurity states with bias $\epsilon$, $\Omega$ is
the tunneling rate, and the $\w_i>0$ and $\lambda_i$ are the frequencies and coupling
constants of the bath oscillators. The bath is completely specified by its propagator at
the ``impurity'' location
\begin{equation}
\label{gw}
    \Gamma(\w)= \sum_{i} \frac{\lambda_{i}^2}{\omega+i0^+ -\omega_{i}}
\end{equation}
whose spectral density $J(\w) = -{\rm Im}\,\Gamma(\w)$
is commonly parameterized as
\begin{equation}
  J(\omega) = 2\pi\, \alpha\, \omega_c^{1-s} \, \omega^s\,,~ 0<\omega<\omega_c\,,\ \ \ s>-1
\label{power}
\end{equation}
where the dimensionless parameter $\alpha$ characterizes the dissipation strength, and
$\omega_c$ is a cutoff energy. The value $s=1$ represents the case of ohmic dissipation.
For $0<s\leq1$, the spin-boson model \eqref{sbm} displays a quantum phase transition
(QPT) between a delocalized and a localized phase, reached for small and large $\alpha$,
respectively.\cite{leggett,KM96,BTV,VTB,BLTV,mvrev,karyn,florens10,wong08,rieger,fehske,guo,plenio11}

NRG has enabled detailed investigations of this QPT.
While statistical-mechanics arguments\cite{mvrev,emery,rieger} suggest that this transition is in the same
universality class as the thermal phase transition of the one-dimensional (1d) Ising
model with $1/r^{1+s}$ long-range interactions, initial NRG results\cite{VTB} were in disagreement
with this quantum-to-classical correspondence (QCC). In particular, NRG
delivered $s$-dependent critical exponents which obeyed hyperscaling for all $0<s<1$.
Those exponents agreed with the ones of the Ising model for $s>1/2$, but disagreed for
$s<1/2$ where the Ising model is above its upper-critical dimension and displays
mean-field behavior without hyperscaling.\cite{luijten,fisher}
This apparent violation of QCC prompted further numerical investigations. Quantum Monte Carlo
(QMC)\cite{rieger} and exact-diagonalization\cite{fehske} studies led to the
opposite conclusion, namely that the critical behavior of the spin-boson model for
$s<1/2$ is classical and of mean-field type.

Meanwhile, two different errors of the NRG method have been identified,\cite{erratum}
which can be held responsible for incorrect results regarding critical properties. These
two errors are the Hilbert-space truncation error, arising from the fact that the
infinite Hilbert space of each bath harmonic oscillator needs to be truncated to $N_b$
local states, and the mass-flow error, which is rooted in the iterative diagonalization
procedure of NRG and leads to an effective shift of model parameters along the NRG run.
The mass-flow error has been discussed in some detail in Ref.~\onlinecite{flow} where it
has been shown to give rise to an incorrect exponent $x$, describing the thermal
divergence of the susceptibility at criticality. Ref.~\onlinecite{flow} also implemented
a recipe to correct the mass-flow error which then allowed to recover the mean-field
value $x=1/2$ for $s<1/2$.
The problem of Hilbert-space truncation has been solved in Ref.~\onlinecite{guo}, by
employing an optimized boson basis within a variational matrix-product-state approach on
the Wilson chain. Applied to the spin-boson model with $s<1/2$, this method not only
delivered the expected mean-field values for the order-parameter exponents $\beta$ and
$\delta$ if the variational bosonic Hilbert space is taken to be (effectively) infinite,
but also showed that artificially restricting the Hilbert space causes a crossover at
small energies to the erroneous non-mean-field behavior of NRG.

The shortcomings of NRG have independently become clear in studies of dissipative
anharmonic-oscillator models: These models are in one-to-one correspondence with a 1d
Ising $\phi^4$ theory with $1/r^{1+s}$ interactions and must therefore display a
mean-field transition for $s<1/2$. Nevertheless, standard NRG delivered\cite{flow} the same
non-mean-field exponents as for the spin-boson model.

Together, these developments show that the initial NRG results\cite{VTB} were incorrect
and that the QCC is fulfilled for the sub-ohmic spin-boson model. However, it has remained a
puzzle {\it why and how} the NRG could produce an internally consistent set of exponents which
moreover obeyed hyperscaling. This question is of obvious relevance for further
applications of NRG.
It is the purpose of the present paper to close this gap. We discuss Hilbert-space
truncation in some detail, showing that the combination of Hilbert-space truncation and
mass flows acts like a non-analytic term in the Landau expansion of the free energy. At
criticality, this in turn forces the system to effectively remain locked at the
upper-critical dimension for all $s\leq 1/2$. As a result, a set of ``trivial'' exponents
with hyperscaling properties is obtained.
While Hilbert-space truncation and mass flow may be thought of as independent
sources of error, their effects are intertwined: As we argue below, the non-analytic
term in the Landau expansion can be understood in terms of a {\em truncated mass flow}.

We note that recent papers\cite{si09,kirchner10,kirchner11} insist on the breakdown
of QCC for the spin-boson model. However, the concrete numerical results obtained in
Refs.~\onlinecite{si09,kirchner11} are fully consistent with the scenario proposed here,
as will be discussed below. Moreover, the objections raised against QCC in
Ref.~\onlinecite{kirchner10} do not apply. Hence, we believe the interpretation given in
Refs.~\onlinecite{si09,kirchner10,kirchner11} is incorrect.

\subsection{Outline}

The remainder of the paper is organized as follows:
Sec.~\ref{sec:landau} summarizes the spin-boson model, together with the standard
derivation of its QCC via a Feynman path integral. It then reviews the properties of the
resulting $\phi^4$ theory -- whose language will be heavily used in the paper -- in
particular the mean-field regime of $s<1/2$ where Landau theory applies. Last but not
least, Sec.~\ref{sec:landau} proposes that supplementing the field theory by a
certain non-analytic term yields exactly the non-mean-field exponents seen in NRG -- this
non-analytic term will derived and discussed in the remainder of the paper.
Sec.~\ref{sec:flow} dives into technical details of NRG, by repeating and extending the
discussion of the mass-flow error as given in Ref.~\onlinecite{flow} -- this includes
sub-leading corrections as addressed in Ref.~\onlinecite{kirchner11}.
Finally, Sec.~\ref{sec:trunc} is devoted to the Hilbert-space truncation error and the
truncated mass flow. It will highlight the crossover scales induced by the truncation and
their influence on physical observables.
A general discussion of NRG and its applicability to mean-field critical phenomena closes
the paper.


\section{Spin-boson model and field theory}
\label{sec:landau}

On general grounds, one expects that the quantum phase transition in the spin-boson model
can be described by a field theory for an Ising (i.e. real scalar) order parameter
$\phi(\tau)$, where $\langle \phi \rangle$ corresponds to the local magnetization
$\langle\sigma_z\rangle$.

\subsection{Derivation}
\label{sec:feyn}

Such a field theory can be derived from the Feynman path-integral
representation\cite{negele} of the partition function of ${\cal H}_{\rm SB}$. One starts
with a Trotter decomposition of the imaginary time axis into $N$ intervals of length
$\Delta\tau=\beta/N$. The operators $\exp(-{\cal H}_{\rm SB} \Delta\tau)$ are then
evaluated by inserting identity operators in terms of the eigenstates of $\sigma_z$, with
eigenvalues $S=\pm 1$, and the bath-oscillator positions, $q_i$, for each time slice $k$.
(Utilizing coherent states is not necessary, as ${\cal H}_{\rm SB}$ contains
distinguishable objects only.\cite{coh_foot}) One arrives at:
\begin{eqnarray}
\label{feyn}
\mathcal{S}
= &-& \sum_{k=1}^N [\frac{1}{2} \ln \coth (\Omega\,\Delta\tau)] S_k S_{k+1} \\
&+& \sum_{k=1}^N \sum_i \big[ \lambda_i \sqrt{m_i\w_i} \Delta\tau S_k q_{k,i} \nonumber \\
&&~~~+ \frac{m_i}{2} \frac{(q_{k,i}-q_{k-1,i})^2}{\Delta\tau} + \Delta\tau \frac{m_i}{2} \w_i^2 q_{k,i}^2
\big]
\nonumber
\end{eqnarray}
where $m_i$ are the masses of the bath oscillators, and the limit $\Delta\tau\to 0$ has
to be taken at the end as usual.\cite{negele} Note that the action of $\sigma_x$ in
${\cal H}_{\rm SB}$ has been evaluated directly.
The oscillator bath is best treated in Fourier (i.e. frequency) space. Defining
\begin{equation}
\bar{S}_n = \frac{1}{\sqrt{N}} \sum_k e^{-2\pi i n k/N} S_k
\end{equation}
for integer wavenumbers $n=0,\ldots,N-1$ and using the explicit form of the spectral density
\eqref{power}, one finds for the bath term the exact result:
\begin{eqnarray}
\mathcal{S}_b &=& \sum_n \bar{S}_n \bar{S}_{-n} \, \pi\alpha \,
\frac{(\w_c\Delta\tau)^{1-s}}{\sin(\pi s/2)} \left|2 \sin \frac{\pi n}{N}\right|^s
\end{eqnarray}
The integrals in partition function involving the bath-oscillator coordinates are Gaussian,
such that the bath oscillators can be integrated out exactly.\cite{gauss_foot}
After transforming back to real space (i.e. time) the full action takes the form
\begin{eqnarray}
\label{Sdisc}
\mathcal{S}
&=& - \sum_{k=1}^N [\frac{1}{2} \ln \coth (\Omega\,\Delta\tau)] S_k S_{k+1} \\
  &+& \frac{1}{2} \sum_{kk'} K(k\!-\!k')\, (S_k - S_{k'})^2 .
\nonumber
\end{eqnarray}
The bosonic bath thus generates a long-ranged self-interaction in imaginary
time, with the long-distance behavior
\begin{equation}
K(k\!-\!k') = \alpha \Gamma(1+s) (\w_c\Delta\tau)^{1-s} \frac{1}{(k-k')^{1+s}},
\end{equation}
valid for $1\ll|k-k'|\ll N$. The representation \eqref{Sdisc} is real and free of a
Berry-phase term.\cite{coh_foot}

At this stage, one may decide to interpret the $S_k$ (for fixed finite $\Delta\tau$) as
discrete spins, which form an Ising chain with long-range interactions. Its thermodynamic
limit is obtained as $\beta\to\infty$, and one can show that the long-distance properties
are independent of $\Delta\tau$. (This is somewhat non-trivial, as the nearest-neighbor
coupling in \eqref{Sdisc} diverges as $\Delta\tau\to 0$, while the long-range coupling
vanishes in the same limit.)
Speculations that the limits $\Delta\tau\to0$ and $\beta\to\infty$ might not
commute\cite{VTB} have turned out to be unwarranted, as has been shown by explicit numerical
computations.\cite{rieger}
The phase transition in this Ising model itself is described by a continuum $\phi^4$
theory with long-range interactions.\cite{luijten}

Alternatively, one may depart from Eq.~\eqref{feyn} and relax the hard-length constraint
on the Ising variables. Assuming a slowly varying Ising order-parameter field $\phi$ then
allows to obtain a continuum theory with finite coefficients, which is -- of course --
the same $\phi^4$ theory as obtained for the discrete Ising model.

This $\phi^4$ theory reads
\begin{equation}
\mathcal{S} = \int \frac{d\w}{2\pi} (m_0 + |\w|^s) |\phi(i\w)|^2 + \int d\tau
\left[u \phi^4(\tau) + \bar{\epsilon} \phi(\tau) \right]
\label{phi4}
\end{equation}
the continuous frequency integral has to be replaced by a Matsubara sum for $T>0$. The field
$\phi$ has been re-scaled such that the prefactor of the $|\w|^s$ term equals unity, and
$\bar{\epsilon}\propto\epsilon$ represents the field conjugate to the order parameter.
In Eq.~\eqref{phi4} an increase of the (bare) mass parameter $m_0$ corresponds to a decrease in
the dissipation strength $\alpha$, with the critical point located at $m_0=m_c$ or
$\alpha=\alpha_c$.

By universality arguments, the same field theory describes a dissipative anharmonic oscillator
\begin{eqnarray}
\label{dao}
{\cal H}_{\rm DHO} &=& \Omega a^\dagger a + \frac{\epsilon}{2}(a+a^\dagger)
+ U n_a (n_a-1)
\\
&+& \frac 1 2 \sum_{i} \lambda_{i} (a + a^\dagger) ( b_{i} + b_{i}^{\dagger} ) +
\sum_{i} \w_{i} b_{i}^{\dagger} b_{i} ,
\nonumber
\end{eqnarray}
where $\Omega>0$ is the bare ``impurity'' oscillator frequency, $n_a = a^\dagger a$, and
$U$ parameterizes the anharmonicity.
For $U=\infty$, ${\cal H}_{\rm DHO}$ is equivalent to ${\cal H}_{\rm SB}$, because all
impurity states except $n_a=0,1$ are projected out.
In the opposite limit of small $U$ the interaction term of ${\cal H}_{\rm DHO}$
\eqref{dao} is expected to follow the same RG flow as does $u$ in $\mathcal{S}$
\eqref{phi4}.

\subsection{Critical exponents}

The field theory \eqref{phi4} can be utilized to calculate critical exponents, with the
following definitions:
\begin{eqnarray}
\phi(\alpha > \alpha_c,T=0,\epsilon=0)
&\propto& (\alpha-\alpha_c)^{\beta}, \nonumber\\
\chi(\alpha < \alpha_c,T=0) &\propto& (\alpha_c-\alpha)^{-\gamma},
\nonumber\\[-1.75ex]
\label{exponents} \\[-1.75ex]
\phi(\alpha=\alpha_c,T=0) &\propto& | \epsilon |^{1/\delta}, \nonumber\\
\chi(\alpha=\alpha_c,T) &\propto&
T^{-x}, \nonumber \\
\chi''(\alpha=\alpha_c,T=0,\omega) &\propto&
|\omega|^{-y} {\rm sgn}(\omega), \nonumber
\end{eqnarray}
where $\phi= \langle \phi \rangle$ and $\chi = d\phi/d\epsilon$.
The last equation describes the dynamical scaling of $\chi$. 

The $\phi^4$ theory \eqref{phi4} has been analyzed in detail in
Ref.~\onlinecite{luijten}. Power counting yields the scaling dimensions at criticality:
\begin{eqnarray}
\label{scaldim}
{\rm dim}[\phi(\tau)] &=& (1-s)/2 \,,\\
{\rm dim}[u] &=& 1-4{\rm dim}[\phi(\tau)] = 2s-1 \,,
\nonumber
\end{eqnarray}
i.e., the system is above (below) its upper-critical dimension
for $s<1/2$ ($s>1/2$). Consequently, the transition is controlled by a Gaussian fixed
point for $s<1/2$, with critical exponents
\begin{equation}
\beta_{\rm MF}  = 1/2\,,~
\gamma_{\rm MF}  = 1\,,~
\delta_{\rm MF} = 3\,,~
\nu_{\rm MF} = 1/s\,,
\end{equation}
where the $s$ dependence of the correlation-length exponent $\nu$ reflects the influence
of the long-range interactions. Moreover, the exponents $x$ and $y$, related to the
finite-size-scaling and anomalous decay exponents of the classical Ising model, are
different:
\begin{equation}
x_{\rm MF}  = 1/2\,,~
y_{\rm MF}  = s\,.
\end{equation}
In contrast, for $s>1/2$ the exponents take non-trivial values. Hyperscaling allows to
deduce a few exact results, given that $y=s$ is known to be exact:\cite{fisher,suzuki}
\begin{equation}
x_{\rm hyp} = y_{\rm hyp}  = s\,,~
\delta_{\rm hyp} = \frac{1+s}{1-s}\,.
\end{equation}

\subsection{Temperature-dependent mass and susceptibility}
\label{sec:susc}

As a prerequisite for the following considerations, we now explicitly derive the
temperature dependence of the (renormalized) order-parameter mass at criticality, which
is related to the susceptibility by $1/\chi(T) = m(T)$.

Below the upper-critical dimension, i.e., for $s>1/2$, naive scaling applies. Given the
zero-temperature form of the critical propagator $1/G_\phi \propto \w^s$, the
interaction-induced mass at criticality has to scale as $m(T) \propto T^s$, which yields
$x=s$ as stated above.

In contrast, in the mean-field regime of $s<1/2$, $m(T)$ can be obtained in lowest-order
perturbation theory in $u$. The tadpole diagram at criticality evaluates to
\begin{equation}
\label{sig1}
\Sigma_1 = u T \sum_j \frac{1}{|\w_j|^s + b \w_j^2 + m(T)}
\end{equation}
where $\w_j = 2\pi j T$ are Matsubara frequencies and $b \w_j^2$, representing the
short-range piece of the interaction, has been included for convergence. The
temperature-dependent piece, $\Sigma_1(T)-\Sigma_1(0) = m(T)$, can be obtained by simple
scaling analysis. Assuming $m(T) = a T^x$ we find
\begin{equation}
a T^x = u T^{1-x} \left( \sum_j \frac{1}{|2\pi j|^s T^{s-x} + a} - \int \frac{dj}{|2\pi j|^s T^{s-x}}\right)
\label{tdep}
\end{equation}
For $x>s$ and small $T$ the right-hand side is dominated by the $j=0$ piece of the sum
and hence approaches a constant as $T\to 0$. Demanding the temperature power laws on both
sides of Eq.~\eqref{tdep} to match, we obtain $x=1/2$ as expected\cite{luijten} (which
indeed fulfills $x>s$).

The calculation shows that it is mandatory to include the mass self-consistently into the
internal propagator of the tadpole diagram; this is different from the situation in
$\phi^4$ theories with short-ranged interactions where the tadpole diagram with bare
propagator would be sufficient to determine the shift exponent.

\subsection{Field theory with NRG errors}
\label{sec:serr}

Our central result is that the errors of NRG induce extra non-analytic mass terms in the
order-parameter field theory:
\begin{equation}
\mathcal{S}_{\rm err} =
\int d\tau \, \phi^2(\tau) \left[v T^s + w |\langle\phi\rangle|^{2s/(1-s)}\right]\,,
\label{Serr}
\end{equation}
with $v$ and $w$ depending on parameters of the NRG algorithm.
The first term is a consequence of the mass flow, as discussed in Ref.~\onlinecite{flow}
and summarized in Sec.~\ref{sec:flow} below. It is responsible for the incorrect result
$x=s$ found using NRG for $s<1/2$.
The second term originates from Hilbert-space truncation, with the prefactor $w$ scaling
with the number $N_b$ of bosons as $w\propto N_b^{-s/(1-s)}$. It will be derived in
detail in Sec.~\ref{sec:trunc}.

Let us discuss the effect of the second term on critical exponents. Approximating $\phi(\tau)$
as static in $\mathcal{S}+\mathcal{S_{\rm err}}$ and setting $T=0$ leads to a Landau free
energy
\begin{equation}
F(\phi) = \bar{\epsilon} \phi + m \phi^2 + w |\phi|^{2/(1-s)} + u \phi^4 \,.
\label{Fsing}
\end{equation}
It is important to note that fluctuation corrections to this mean-field picture arise
from the quartic $u$ term only (as $w$ is not an interaction), and hence will influence
the critical behavior only for $s>1/2$, where moreover $2/(1-s) > 4$ and hence the $w$
term is subleading.
In contrast, for $s<1/2$ simply minimizing $F(\phi)$ \eqref{Fsing} is sufficient to calculate the
zero-temperature exponents. For $w=0$ this of course results in $\beta_{\rm MF} = 1/2$
and $\delta_{\rm MF}=3$. However, a non-zero $w$ term dominates over the
quartic $u$ term for $s<1/2$, and minimizing $F(\phi)$ then gives
\begin{equation}
\label{betanrg}
\beta = \frac{1-s}{2s}\,,~
\delta=\frac{1+s}{1-s}\,,
\end{equation}
which are exactly the exponents observed\cite{VTB} in NRG!
The equation of state derived from \eqref{Fsing} is also consistent with the
comprehensive scaling analysis of NRG results at different $N_b$ performed in
Ref.~\onlinecite{tong11}.

In the remainder of the paper, we discuss the errors of the bosonic NRG
and the consequences of Eq.~\eqref{Serr} in detail.


\section{NRG and mass flow}
\label{sec:flow}

The mass-flow problem of the bosonic NRG has been investigated in
Ref.~\onlinecite{flow}. Here we shall summarize the discussion and also provide an
analysis of certain subleading corrections.

\begin{figure}[!t]
\includegraphics[width=2.9in,clip]{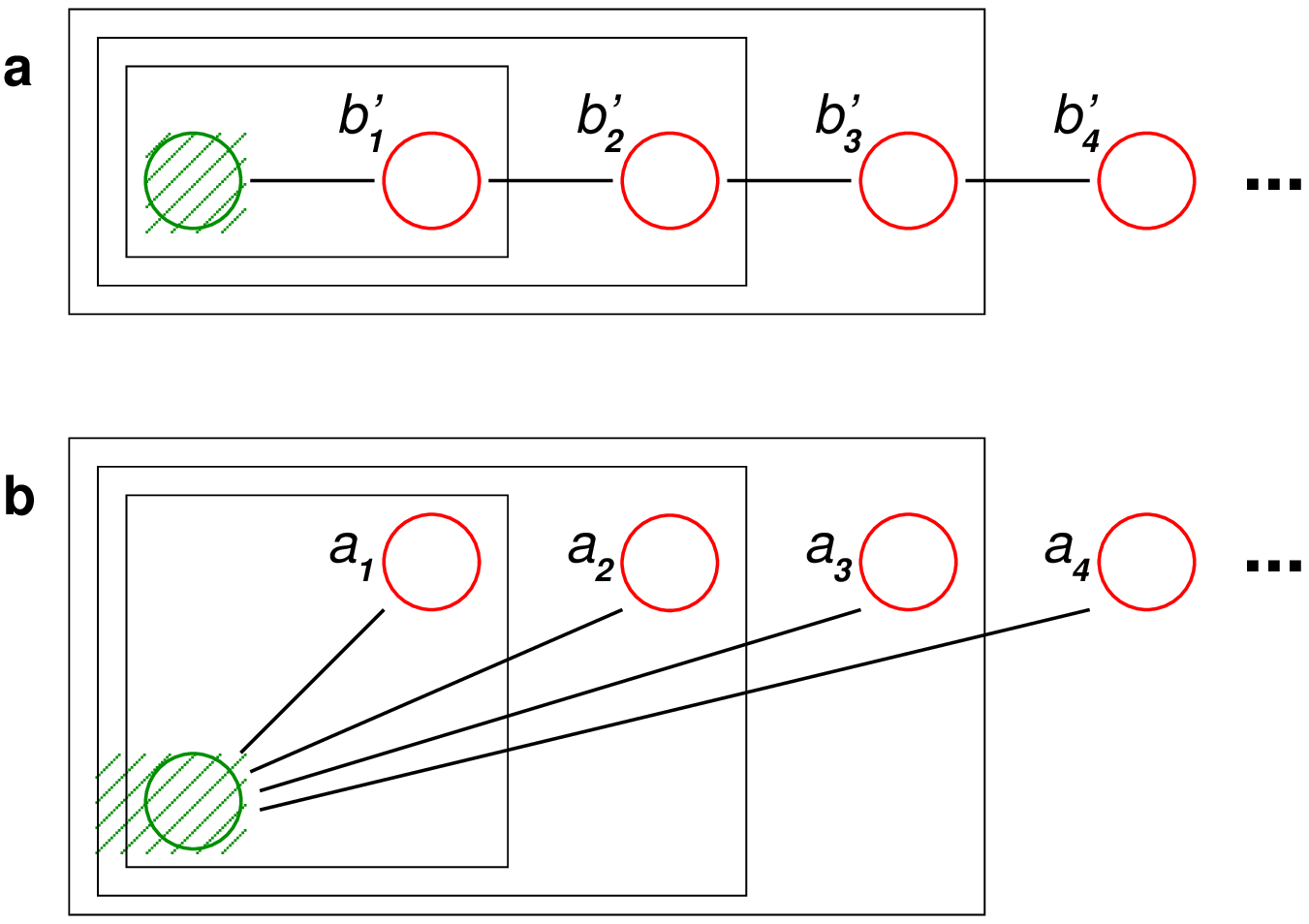}
\caption{
(Color online)
Structure of the NRG Hamiltonian, with the bath represented by a semi-infinite Wilson
chain, with bath operators $b'_n$. The boxes indicate the iterative diagonalization
scheme.
}
\label{fig:itdiag}
\end{figure}

Within the NRG algorithm, the bath is represented by a semi-infinite (``Wilson'') chain,
Fig.~\ref{fig:itdiag}, such that the impurity is coupled to the first site of the chain
only, and the local density of states at this first site is a discrete approximation to
the bath density of states.\cite{wilson75,nrgrev} Due to the logarithmic discretization,
the site energies $\varepsilon_n$ and hopping matrix elements $t_n$ decay exponentially
along the chain according to $\w_c \Lambda^{-n+1}$, where $\Lambda$ is the discretization
parameter.
We define a Hamiltonian $\mathcal{H}_n$ for the impurity plus the first $n$ sites of the
Wilson chain, and we denote by $\Gamma_n(\w)$ the corresponding propagator at the
impurity site of this $n$-site bath. Then, $\mathcal{H}_\infty$ with bath
$\Gamma_\infty(\w)$ is the discretized version of the original problem.

NRG proceeds by iteratively diagonalizing the Hamiltonian: First, $\mathcal{H}_1$ is
diagonalized and the lowest $N_s$ eigenstates are kept. Then, the next bath site is added
to form $\mathcal{H}_2$, the new system is diagonalized, and again the lowest $N_s$
eigenstates are kept (which are approximations to the lowest states of $\mathcal{H}_2$).
As the characteristic energy scale of the low-lying part of the eigenvalue spectrum
decreases by a factor of $\Lambda$ in each step, this process is repeated until the
desired lowest energy is reached.
Thermodynamic observables at a temperature $T_n = \w_c \Lambda^{-n+1} / \bar\beta$
are typically calculated via a thermal average taken from the eigenstates at NRG step $n$.
Here, $\bar\beta$ is a parameter of order unity which is often chosen as
$\bar\beta=1$.

\subsection{Mass-flow error}

The iterative diagonalization procedure implies that, at NRG step $n$, the chain sites
$n+1$, $n+2$, \ldots have not yet been taken into account, i.e., the effect of those
sites does not enter thermodynamic observables at temperature $T_n$. The
``missing'' sites imply a missing contribution to the real part of the bath propagator. As
detailed in Ref.~\onlinecite{flow}, the zero-frequency value of this missing real part,
${\rm Re}(\Gamma_\infty-\Gamma_n)(\w\!=\!0)$, scales as $(\bar\beta T_n)^s$ for a power-law bath
spectrum, Eq.~\eqref{power}.
In the field-theory language, the missing real part implies a temperature-dependent
variation of the order-parameter mass. This is most clearly seen for the dissipative
oscillator model in Eq.~\eqref{dao} where ${\rm Re}\,\Gamma(\w=0)$ directly renormalizes
the oscillator frequency, $\Omega \rightarrow \Omega + {\rm Re} \Gamma(\w\!=\!0)$ at
$U=0$ (note that ${\rm Re}\,\Gamma(\w\!=\!0)<0$).
Similarly, for the spin-boson model it is clear that the missing real part effectively
leads to a temperature-dependent variation of model parameters. Provided that the spin-boson
model renormalizes to the same $\phi^4$ field theory as the dissipative oscillator, as
shown in Sec.~\ref{feyn}, it is clear that this variation is in the $\phi^4$ mass.

Thus, the mass-flow error generates an artificial mass contribution $m_a = v T^s$ -- this
is the first term in $\mathcal{S}_{\rm err}$ \eqref{Serr}. The prefactor $v$ scales as
$\bar\beta^s$, but also depends on the model and parameter regime under consideration
(note that the mapping of the microscopic $\mathcal{H}$ to the field theory involves
coarse graining). Moreover, interaction effects will generate subleading contributions to
$m_a$, see below.

\subsection{Temperature dependence of $\chi$ in the presence of mass flow}

As derived in Sec.~\ref{sec:susc}, the irrelevant interaction $u$ leads to an
order-parameter mass $\propto T^{1/2}$ along the flow towards the Gaussian fixed point
for $s<1/2$. Consequently, the physical susceptibility follows $\chi \propto T^{-1/2}$
(Ref.~\onlinecite{luijten}).
However, the artificial mass $m_a \propto T^s$ dominates the physical mass at low $T$,
leading to the unphysical result $1/\chi \propto T^{s}$ -- which happens to coincide with the
physical result for an interacting fixed point with hyperscaling.

Considering the results in Ref.~\onlinecite{kirchner11}, it is worth evaluating the
subleading correction to the mass-flow-determined $m_a = vT^{s}$. This piece,
$m_2(T)$, is now arising from the quartic interaction and can be evaluated in
lowest-order perturbation theory. In analogy to Eq.~\eqref{sig1} we need to evaluate
$\Sigma_1(T)-\Sigma_1(0) = m_2(T)$, with
\begin{equation}
\Sigma_1 = u T \sum_n \frac{1}{|\w_n|^s + b \w_n^2 + v T^s + m_2(T)}
\end{equation}
where the propagator now contains {\em both} the artificial and the interaction-generated
mass contribution. Assuming $m_2(T) = v_2 T^{x_2}$ with $x_2>s$, the $m_2$ term in the
internal propagator is subleading. Consequently, the equation determining $m_2$ reads
\begin{equation}
v_2 T^{x_2} = u T \left( \sum_j \frac{1}{(|2\pi j|^s +v) T^s} - \int \frac{dj}{|2\pi j|^s T^s}\right)
\label{tdep2}
\end{equation}
and one immediately arrives at $x_2 = 1-s$. Hence, the
presence of the artifical mass modifies the interaction-generated mass.\cite{sub_foot}
Indeed, the NRG results for $\chi(T)$ for both the spin-boson model\cite{kirchner11} and
the dissipative anharmonic oscillator\cite{flow} are consistent with $x_2 = 1-s$ and thus
in line with the above calculation. The calculation falsifies the claim of
Ref.~\onlinecite{kirchner11} that $x_2>1/2$ would be incompatible with the mass-flow
scenario.

\subsection{Mass-flow correction}

Ref.~\onlinecite{flow} proposed an empirical algorithm to cure the mass-flow problem.
This is based on compensating the NRG-induced mass flow by including an explicit
temperature-dependent term, with coefficient $\kappa T^s$, in the Hamiltonian which
shifts the critical point similar to the effect of ${\rm Re}\,\Gamma$. For the spin-boson
model, the natural choice is to add a temperature-dependent piece to the tunneling matrix
element $\Omega$. In this procedure, $\kappa$ remains a free parameter, and Ref.~\onlinecite{flow}
proposed to adjust its value at criticality such that the unphysical $T^s$ piece in
$1/\chi$ is removed.
This algorithm, dubbed NRG$^\ast$, was shown to work well and to yield the expected
mean-field behavior $1/\chi\propto T^{1/2}$ at low temperature for $s<1/2$ whereas the
$T^s$ behavior was stable for $s>1/2$. Therefore, the QCC for the exponent $x$ has been
confirmed by NRG$^\ast$.

Ref.~\onlinecite{kirchner11} recently questioned this result, based on the observation
that $1/\chi(T)$ from NRG$^\ast$, which follows $T^{1/2}$ at low $T$, displays an unusual
{\em downward}-deviation from this law at higher $T$ (see Figs.~7b and 8b of
Ref.~\onlinecite{flow}). Ref.~\onlinecite{kirchner11} claimed that this would reflect an
underlying temperature power law with exponent smaller than $1/2$, then supposedly
inconsistent with mean-field behavior.
Here we argue instead that the downward-deviation can be straightforwardly
understood in terms of the mass flow and its compensation. As explained in
Ref.~\onlinecite{flow}, the mass-flow correction term with fixed $\kappa$ only
compensates for the mass flow near a specific fixed point, and different fixed points
would require different $\kappa$. In other words, if the proper $\kappa_0$ for the
low-temperature fixed point has been chosen, then this will compensate the artificial
mass only to leading order, but a subleading term will be left over.
For the spin-boson model, one can easily understand the sign of the subleading term by
noting that NRG$^\ast$ produces an overcompensation of the mass flow at elevated
temperatures or energies: At high energies, ${\rm Re}\,\Gamma$ influences the mass very
little (note that it couples to $\sigma_z^2$ in Eq.~\eqref{sbm} in the short-time limit).
Then, at high energies, only the mass-flow correction is active, which leads to a
downward correction of the mass. It is this downward correction that causes the downturn
in $1/\chi(T)$. (We also note that the downturn does not correspond to a new power-law
regime, contrary to the assertion of Ref.~\onlinecite{kirchner11}.)
We conclude that the downturn in $1/\chi(T)$ is not physical, but instead
related to the shortcomings of the simple mass-flow correction in NRG$^\ast$.
The same downturn is observed when applying NRG$^\ast$ to the dissipative anharmonic
oscillator with large $U$.


\section{NRG and Hilbert-space truncation}
\label{sec:trunc}

In this section, we discuss how the truncation of the bosonic Hilbert space at each site
of the Wilson chain, i.e., using only the lowest $N_b$ occupation-number eigenstates of
the oscillator as basis, influences the evaluation of observables in the spin-boson model.
In particular, we will explain how the non-analytic $w$ term in the effective field theory
\eqref{Serr} emerges.

Hilbert-space truncation is bound to become important if $\sigma_z$ develops a finite
expectation value, either inside the ordered phase or in an applied field $\epsilon$.
For a finite $\langle\sigma_z\rangle$, the bath oscillators will have a finite
displacement $\langle x_n \rangle =  \langle b_n + b_n^\dagger \rangle$.
By considering the system at the localized fixed point, $\Delta=0$, it is easy to show
that -- due to the logarithmic discretization scheme of NRG -- $\langle x_n \rangle$
diverges in the low-energy limit for $s<1$ according to\cite{BLTV}
$\varepsilon_n^{(s-1)/2}$. This implies that the average boson numbers diverge as well:
\begin{equation}
\langle b_n^\dagger b_n \rangle \propto \langle\sigma_z\rangle^2 \varepsilon_n^{s-1} \,.
\label{div}
\end{equation}
Importantly, $\langle\sigma_z\rangle$ in this equation is a temperature-dependent
quantity which thus depends on $n$ as well, but will typically saturate below a
temperature $\Tsat$.

\subsection{Truncated mass flow}

Trivially, for $N_b=\infty$ there are no truncation effects, and $\langle b_n^\dagger b_n
\rangle$ diverges in the low-energy limit $n\to\infty$. For any finite $N_b$, the
divergence of $\langle b_n^\dagger b_n \rangle$ is cut-off at some $\ncut$, where
$\langle b_n^\dagger b_n \rangle$ reaches a value of order $\mathcal{O}(N_b)$. In the NRG
algorithm, $\ncut$ corresponds to a cut-off energy or temperature given by $\Tcut =
\varepsilon_{\ncut}$, below which truncation is relevant. From Eq.~\eqref{div} we deduce $N_b
\propto \langle\sigma_z\rangle^2\,\Tcut^{s-1}$ which yields
\begin{equation}
\label{tcut}
\Tcut \propto N_b^{-1/(1-s)} \, |\langle\sigma_z\rangle|^{2/(1-s)}\,.
\end{equation}

What happens below $\Tcut$? The sites of the Wilson chain with $n>\ncut$ no longer behave
as free bosonic states with energy $\varepsilon_n$: The low-energy states of the shifted
oscillators are missing, because the truncated basis of the lowest $N_b$
occupation-number eigenstates is too small to accommodate them. Qualitatively, the
energies of the oscillators with $n>\ncut$ are systematically shifted upwards. By
Kramers-Kronig relations this implies that the corresponding real part, ${\rm
Re}\,\Gamma_n$, is modified into ${\rm Re}\,\tilde\Gamma_n$ for $n>\ncut$, with ${\rm
Re}\,\tilde\Gamma_\infty$ not reaching ${\rm Re}\,\Gamma_\infty$.

This is where the mass-flow effect comes into play. The artificial mass $m_a(T)$, given by
${\rm Re}(\Gamma_\infty - \Gamma_n)$ from the ``missing'' part of the Wilson chain, will
no longer decrease $\propto T^s$ below $\Tcut$, but will instead approach a constant
value $\mcut$ dictated by ${\rm Re}(\Gamma_\infty-\tilde\Gamma_\infty)$. This constant can be
easily estimated: Assuming that the oscillators with $n>\ncut$ are shifted to infinite
energy, it is given by $\mcut=v\Tcut^s$; corrections to this simple estimate will modify
the prefactor, but not the power-law dependence on $\Tcut$.
These considerations imply that the artificial mass $m_a$ will only flow very little
below $\Tcut$, i.e., the mass flow is truncated at $\Tcut$.

Taking the mass-flow and truncated mass-flow contributions together, we arrive at an
artificial mass term in the order-parameter theory, which scales as $[\max(T,\Tcut)]^s$.
Accounting for different prefactors and inserting Eq.~\eqref{tcut} we arrive at
the non-analytic mass terms as postulated in Sec.~\ref{sec:serr} above:
\begin{equation}
\mathcal{S}_{\rm err} =
\int d\tau \, \phi^2(\tau) \left[v T^s + w |\langle\phi\rangle|^{2s/(1-s)}\right]\,,
\label{Serr2}
\end{equation}
where $w\propto N_b^{-s/(1-s)}$ vanishes as $N_b\to\infty$.

As mentioned in Sec.~\ref{sec:serr} the non-analytic $w$ term asymptotically dominates
over the quartic $u$ term for $s<1/2$ in the mean-field equation of state,
Eq.~\eqref{Fsing}, such that it leads to the incorrect critical exponents \eqref{betanrg}
observed in NRG.

In the power-counting sense, the non-analytic $w$ term is {\em marginal} for all $s$:
Using Eq.~\eqref{scaldim} we have
\begin{equation}
{\rm dim}[w] = 1-\frac{2}{1-s}{\rm dim}[\phi(\tau)] = 0\,.
\end{equation}
The system is thus effectively locked at an upper-critical dimension. This explains
why the NRG exponents \eqref{betanrg} obey hyperscaling, but are nevertheless ``trivial''
in the sense that only simple fractions of $s$ occur. (Note that logarithmic corrections
do not occur, due to the absence of fluctuation effects from $w$.)

\subsection{Example: Order-parameter exponent $\delta$}

To illustrate the above considerations, we now discuss explicitly the behavior of the
magnetization at the critical point in a small applied field $\epsilon$ as function of
temperature $T$, which yields the critical exponent $\delta$ according to $\epsilon
\propto \phi^\delta$ as $T\to 0$ (where $\phi=\langle\sigma_z\rangle$).

We discuss the physical behavior of the $\phi^4$ model \eqref{phi4} in the absence of
systematic errors first. The leading temperature dependence of the magnetization stems
from the temperature-induced order-parameter mass $m(T)$. In the mean-field regime of
$s<1/2$, this can be estimated from the equation of state:
\begin{equation}
0 = \epsilon + m(T) \phi + 3 u \phi^3.
\label{eqstate}
\end{equation}
Since $m(T)\to 0$ as $T\to 0$, $\phi$ will saturate below $\Tsat$, with  $m(\Tsat)
\propto \phi^2(T=0)$. Using $m(T) \propto T^{1/2}$ and $\delta=3$, valid in the
mean-field regime, we obtain
\begin{equation}
\label{tsat1}
\Tsat \sim \epsilon^{4/3}~~~(s<1/2).
\end{equation}
This argument can be generalized to the non-mean-field regime of $s>1/2$, using
$m(T) \propto T^s$, leading to
\begin{equation}
\Tsat \sim \epsilon^{2/(s\delta)}~~(s>1/2).
\label{tsat2}
\end{equation}
From Eq.~\eqref{eqstate} one also sees that, above $\Tsat$, $\phi$ will follow
$\phi \propto \epsilon/m(T)$.

Now we turn to the influence of the systematic NRG errors. This requires to discuss the
relation between $\Tsat$, the physical saturation temperature of $\phi$, and $\Tcut$,
the temperature below which Hilbert-space truncation spoils the calculation.

First assume that $\Tsat>\Tcut$. Then the magnetization will have taken its $T=0$ value
$\phi \propto \epsilon^{1/\delta}$ already above $\Tcut$. From Eq.~\eqref{tcut} we deduce
\begin{equation}
\Tcut \propto \epsilon^{2/[(1-s)\delta]}~~(\mbox{if}~\Tsat>\Tcut).
\end{equation}
Comparing this with Eq.~\eqref{tsat2}, we see
that, for small $\epsilon$, indeed $\Tsat>\Tcut$ consistently holds for $s>1/2$. In such
a case, we expect that the Hilbert-space truncation will not significantly influence the
$T=0$ value of $\phi$. This is in contrast to $s<1/2$, where the assumption $\Tsat>\Tcut$
is found to be violated.

\begin{figure}[!t]
\includegraphics[width=3.4in,clip]{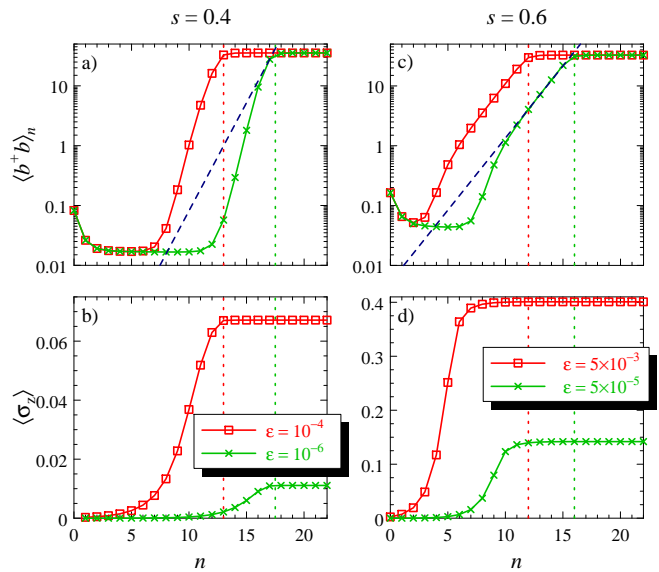}
\caption{ (Color online)
Evolution of the boson occupation number, $\langle b_n^\dagger b_n \rangle$, and the
magnetization, $\langle\sigma_z\rangle$, in the spin-boson model at criticality in an
applied field $\epsilon$, as calculated along the NRG run with $\Delta=1$,
$\Lambda=4$ and $N_b=50$.
a,b) For $s=0.4$ ($\alpha_c=0.0355465$), both quantities saturate at the same $n$, or
equivalently, the same temperature $\Tcut$ (dotted vertical lines), illustrating that
Hilbert-space truncation affects the magnetization.
c,d) In contrast, for $s=0.6$ ($\alpha_c=0.06702411$) the magnetization saturates before $\Tcut$ is reached, and is
thus unaffected by truncation.
The dark dashed lines indicate a divergence $\langle b_n^\dagger b_n \rangle \propto
\varepsilon_n^{s-1}$ according to Eq.~\eqref{div} -- this is not realized for $s=0.4$ because
$\langle\sigma_z\rangle$ is never constant above $\Tcut$, but applies to $s=0.6$ in the
interval $\Tcut<T<\Tsat$ where $\langle\sigma_z\rangle$ is constant.
}
\label{fig:mag}
\end{figure}

Now let us discuss the case $\Tsat<\Tcut$, applying to $s<1/2$, where Hilbert-space
truncation will affect the impurity magnetization. First, note that the estimate of
$\Tsat$ \eqref{tsat1} needs to be modified due to the mass-flow error:
$m(T) \propto T^s$ yields
\begin{equation}
\label{tsat3}
\Tsat \sim \epsilon^{2/(3s)}~~~(s<1/2,~\mbox{mass~flow}).
\end{equation}
At $\Tcut$, the magnetization can be estimated as $\phi \propto \epsilon/m(\Tcut)$. Again
using $m(T) \propto T^s$ and combining this with Eq.~\eqref{tcut} we have
\begin{equation}
\label{ptcut}
\phi(\Tcut) \propto \frac{\epsilon}{\phi(\Tcut)^{2s/(1-s)}}
\end{equation}
and
\begin{equation}
\label{tcut3}
\Tcut \propto \epsilon^{2/(1+s)}~~~(s<1/2,~\mbox{mass~flow}).
\end{equation}
As argued above, due to the truncated mass flow, $m$ remains of order $\Tcut^s$ for all
$T<\Tcut$. Therefore Eq.~\eqref{ptcut} implies that
\begin{equation}
\phi \sim \phi(\Tcut) \propto \epsilon^{(1-s)/(1+s)}
\end{equation}
holds down to $T=0$, resulting in the critical exponent $\delta=(1+s)/(1-s)$ as indeed
observed.

The truncated mass flow is shown using NRG data in Fig.~\ref{fig:mag}. For $s<1/2$
the magnetization $\phi=\langle\sigma_z\rangle$ indeed increases down to $\Tcut$, but
stops to increase below $\Tcut$ -- this proves both $\Tsat<\Tcut$ and $m(T)\approx {\rm
const}$ for $T<\Tcut$. The latter reflects the advertised truncated mass flow and leads
to incorrect scaling.
In contrast, for $s>1/2$ the magnetization saturates already above
$\Tcut$, indicating $\Tsat>\Tcut$ as argued above.

Importantly, Fig.~\ref{fig:mag} illustrates the qualitative difference between $s<1/2$
and $s>1/2$ independent of the knowledge of the underlying field theory:
Hilbert-space truncation limits the magnetization only for $s<1/2$, where it consequently
spoils the physical critical behavior.

\begin{figure}[!t]
\includegraphics[width=2.9in,clip]{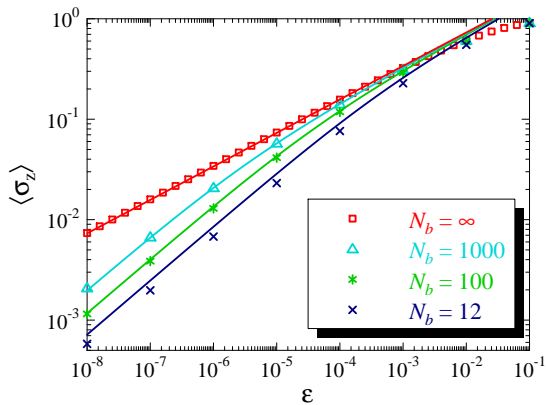}
\caption{ (Color online)
Effect of Hilbert-space truncation on the evaluation of the critical exponent $\delta$ in
the spin-boson model for $s=0.3$. The data points have been obtained using the
variational matrix-product-state approach of Ref.~\onlinecite{guo} with parameters
$\Lambda=2$, $\Delta=0.1$ where $\alpha_c=0.0346142$; for small $N_b$ results from
standard NRG are identical up to minimal deviations.
The lines show the crossover behavior according to the equation of state \eqref{state} with
$\bar{w}=1/50$ and $3u=1/39$ -- this gives an excellent description of the data for
not too small $N_b$.
}
\label{fig:delta}
\end{figure}

Upon increasing the applied field $\epsilon$ for $s<1/2$, $\Tsat$ \eqref{tsat3} will
increase faster than $\Tcut$ \eqref{tcut3}, such that the correct mean-field behavior is
recovered at $T=0$ for intermediate values of $\epsilon$.\cite{erratum}
This crossover is fully described by the equation of state
\begin{equation}
|\bar\epsilon| = \bar{w} N_b^{-s/(1-s)} \phi^{(1+s)/(1-s)} + 3 u \phi^3
\label{state}
\end{equation}
obtained from minimizing \eqref{Fsing} for $m=0$, i.e., at criticality, and setting $w =
\bar{w} N_b^{-s/(1-s)} (1-s)/2$. A comparison of this prediction with numerical data in
shown in Fig.~\ref{fig:delta}, with excellent agreement except for very small $N_b$ where
the continuum approximation w.r.t. $N_b$ is not yet accurate.


\section{Discussion}

We have shown that mass flow, i.e., the fact that the order-parameter mass depends on the
length $n$ of the Wilson chain, is the source of all incorrect exponents obtained by NRG for
the spin-boson model:
(i) The incorrect finite-temperature exponent $x$ arises because the rest of
the chain is discarded when calculating finite-temperature observables at a temperature $T_n$.
(ii) The incorrect zero-temperature exponents $\beta$ and $\delta$ arise because the
chain is effectively truncated at some magnetization-dependent $\ncut$ where the
diverging oscillator shifts can no longer be accomodated within the finite Hilbert space
for each boson site.
In the order-parameter field theory, (i) and (ii) can be captured as artificial
non-analytic mass contributions, Eq.~\eqref{Serr}.

The fact that the resulting NRG exponents are internally consistent and obey hyperscaling
can be rationalized by realizing that both errors of bosonic NRG can be associated with
{\em stationarity} along the NRG flow.
(A) The mass-flow error enforces the temperature-dependent renormalized mass to be stationary
at criticality (which corresponds to naive scaling $m_a\propto T^s$).
(B) The Hilbert-space truncation error cuts the divergence of the boson occupation
number, making it stationary.
Thus, both errors force the NRG flow to show fixed-point behavior -- in the NRG sense of
stationarity of the many-body energies and states along the flow -- in a situation where
the correct physical RG flow does not show stationarity in the same sense. This applies
both to the localized phase of the spin-boson model, where the boson number diverges along the
Wilson chain, and to the flow towards the Gaussian critical point, where the true many-body level
spacing goes to zero as $T\to 0$ because of the irrelevance of the quartic
interaction.\cite{flow}

This finally solves the puzzle: NRG produces incorrect exponents obeying hyperscaling
because its algorithmic deficiences force the NRG flow to be stationary, which implies
naive scaling.
One concludes that it is difficult with traditional NRG to describe critical phenomena
which violate hyperscaling and are controlled by a dangerously irrelevant variable because
in such a case the NRG flow will not be stationary near the physical RG fixed point\cite{pgk_foot} --
this has been (partially) illustrated in Ref.~\onlinecite{flow}. Algorithmic
modifications, e.g. the variational matrix-product-state approach of
Ref.~\onlinecite{guo}, can in principle solve the problem.

The present insights show that NRG-based claims of violated QCC in other Ising-symmetric
impurity models with sub-ohmic bosonic baths\cite{kevin,glossop09} should be revisited.
Also, implications for the calculation of dynamical
properties\cite{dyn1,dyn2,dyn3,florens11} need to be investigated.
In bosonic impurity models with higher symmetries, QCC was found to be fulfilled for a
rotor model,\cite{rotor} but argued to be violated for XY-symmetric spin-boson\cite{guo}
and SU(2)-symmetric Bose-Kondo models.\cite{kisi} These results suggest that the
violation of QCC is related to the presence of multiple baths (or long-ranged
interactions) which couple to different {\em non-commuting} impurity operators. Clearly,
a deeper understanding of the relevant non-classical phase transitions is desirable.


\acknowledgments

We thank J. von Delft, K. Ingersent, S. Kirchner, Q. Si, N. Tong, and, in particular, S.
Florens, R. Narayanan, and T. Voj\-ta for discussions, and R. Bulla, J. von Delft, C.
Guo, H. Rieger, N. Tong, and A. Weichselbaum for collaborations on related topics. We
also thank C. Guo for providing the data in Fig.~\ref{fig:delta}.
This research has been supported by the DFG (FOR 960) and the GIF.




\begin{thebibliography}{99}

\bibitem{wilson75}
K. G. Wilson,
Rev. Mod. Phys. {\bf 47}, 773 (1975).

\bibitem{nrgrev}
R. Bulla, T. Costi, and T. Pruschke,
Rev. Mod. Phys. {\bf 80}, 395 (2008).

\bibitem{BTV}
R. Bulla, N. Tong, and M. Vojta,
Phys. Rev. Lett. {\bf 91}, 170601 (2003).

\bibitem{BLTV}
R. Bulla, H.-J. Lee, N. Tong, and M. Vojta,
Phys. Rev. B {\bf 71}, 045122 (2005).

\bibitem{leggett}
A.~J. Leggett, S. Chakravarty, A.T. Dorsey, M.P.A. Fisher, A. Garg, and W. Zwerger,
Rev. Mod. Phys. {\bf 59}, 1 (1987).

\bibitem{KM96}
S. Kehrein and A. Mielke, Phys. Lett. A {\bf 219}, 313 (1996).

\bibitem{VTB}
M. Vojta, N. Tong, and R. Bulla,
Phys. Rev. Lett. {\bf 94}, 070604 (2005).

\bibitem{mvrev}
M. Vojta, Phil. Mag. {\bf 86}, 1807 (2006).

\bibitem{karyn}
K. Le Hur, P. Doucet-Beaupre, and W. Hofstetter,
Phys. Rev. Lett. {\bf 99}, 126801 (2007).

\bibitem{rieger}
A. Winter, H. Rieger, M. Vojta, and R. Bulla,
Phys. Rev. Lett. {\bf 102}, 030601 (2009).

\bibitem{fehske}
A. Alvermann and H. Fehske,
Phys. Rev. Lett. {\bf 102}, 150601 (2009).

\bibitem{wong08}
H. Wong and Z.-D. Chen,
Phys. Rev. B {\bf 77}, 174305 (2008).

\bibitem{guo}
C. Guo, A. Weichselbaum, J. von Delft, and M. Vojta,
preprint arXiv:1110.6314.

\bibitem{florens10}
S. Florens, D. Venturelli, and R. Narayanan,
Lect. Notes Phys. {\bf 802}, 145 (2010).

\bibitem{plenio11}
A. W. Chin, J. Prior, S. F. Huelga, and M. B. Plenio,
Phys. Rev. Lett. {\bf 107}, 160601 (2011).

\bibitem{emery}
V. J. Emery and A. Luther, \prb {\bf 9}, 215 (1974).

\bibitem{luijten}
E. Luijten and H. W. J. Bl\"ote,
\prb {\bf 56}, 8945 (1997).

\bibitem{fisher}
M. E. Fisher, S.-k. Ma, and B. G. Nickel,
\prl {\bf 29}, 917 (1972).

\bibitem{erratum}
M. Vojta, N. Tong, and R. Bulla,
Phys. Rev. Lett. {\bf 102}, 294904(E) (2009).

\bibitem{flow}
M. Voj\-ta, R. Bulla, F. G\"uttge, and F. B. Anders,
Phys. Rev. B {\bf 81}, 075122 (2010).

\bibitem{si09}
S. Kirchner, Q. Si, and K. Ingersent,
Phys. Rev. Lett. {\bf 102}, 166405 (2009).

\bibitem{kirchner10}
S. Kirchner,
J. Low Temp. Phys. {\bf 161}, 282 (2010).

\bibitem{kirchner11}
S. Kirchner, Q. Si, and K. Ingersent,
preprint arXiv:1111.6623.

\bibitem{negele}
J. W. Negele and H. Orland,
{\it Quantum Many-Particle Systems},
Addison-Wesley (Reading), 1988.

\bibitem{coh_foot}
A representation of ${\cal H}_{\rm SB}$ using coherent spin states, and consequently an
imaginary Berry phase, is possible as well.\cite{kirchner10} However, this by itself does
not imply the invalidity of QCC, as both this coherent-state representation and the real
representation of Sec.~\ref{sec:feyn} faithfully represent the physics of the quantum
model.

\bibitem{gauss_foot}
None of the objections in Sec. 2 of Ref.~\onlinecite{kirchner10} applies.

\bibitem{suzuki}
M. Suzuki, Prog. Theor. Phys. {\bf 49}, 424, 1106, 1440 (1973).

\bibitem{tong11}
Y.-H. Hou and N.-H. Tong,
Eur. Phys. J. B {\bf 78}, 127 (2010);
N.-H. Tong and Y.-H. Hou,
preprint arXiv:1012.5615.

\bibitem{sub_foot}
Ref.~\onlinecite{flow} asserted that, in the presence of mass flow, the temperature
dependence of the order-parameter mass for $s<1/2$ follows $v T^s + v_2 T^{1/2}$, with the
second term representing the interaction-generated physical mass. This
assertion was incorrect: As detailed in Sec.~\ref{sec:flow}, the interaction-generated
mass is modified by the artificial mass, such that the correct behavior is $v T^s + v_2
T^{1-s}$.

\bibitem{pgk_foot}
The fermionic pseudogap Kondo model with bath exponent $r>1$ provides an example for a
phase transition above an upper-critical dimension, with hyperscaling violated, where NRG
nevertheless works.\cite{GBI} This is related to the fact that the effective ``Gaussian'' theory
is that of a level crossing, without dangerously irrelevant variables:\cite{MVLF}
The many-body level spectrum has a well-defined finite spacing even at the ``Gaussian'' fixed
point.

\bibitem{GBI}
C.~Gon\-za\-lez-Buxton and K.~Ingersent,
Phys. Rev. B {\bf 57}, 14254 (1998).

\bibitem{MVLF}
M. Vojta and L. Fritz,
Phys. Rev. B {\bf 70}, 094502 (2004);
L. Fritz and M. Vojta,
{\em ibid.} {\bf 70}, 214427 (2004).

\bibitem{kevin}
M. T. Glossop and K. Ingersent,
Phys. Rev. Lett. {\bf 95}, 067202 (2005);
Phys. Rev. B {\bf 75}, 104410 (2007).

\bibitem{glossop09}
M. Cheng, M. T. Glossop, and K. Ingersent,
Phys. Rev. B {\bf 80}, 165113 (2009).


\bibitem{dyn1}
A. Chin and M. Turlakov,
Phys. Rev. B {\bf 73}, 075311 (2006).

\bibitem{dyn2}
F. B. Anders, R. Bulla, and M. Vojta,
Phys. Rev. Lett. {\bf 98}, 210402 (2007).

\bibitem{dyn3}
P. Nalbach and M. Thorwart,
Phys. Rev. B {\bf 81}, 054308 (2010).

\bibitem{florens11}
S. Florens, A. Freyn, D. Venturelli, and R. Narayanan,
Phys. Rev. B {\bf 84}, 155110 (2011).

\bibitem{rotor}
M. Al-Ali and T. Vojta,
Phys. Rev. B {\bf 84}, 195136 (2011).

\bibitem{kisi}
L. Zhu, S. Kirchner, Q. Si, and A. Georges,
Phys. Rev. Lett. {\bf 93}, 267201 (2004);
S. Kirchner and Q. Si,
preprint arXiv:0808.2647.


\end{thebibliography}
\end{document}